# Comparative Design-Based Research: How Afterschool Programs Impact Learners' Engagement with a Video Game Codesign

Jaemarie Solyst, Carnegie Mellon University, jsolyst@andrew.cmu.edu
Judith Odili Uchidiuno, Georgia Institute of Technology, jiou3@gatech.edu
Erik Harpstead, Carnegie Mellon University, eharpste@andrew.cmu.edu
Jonaya Kemper, Carnegie Mellon University, jkemper@andrew.cmu.edu
Ross Higashi, Carnegie Mellon University, rhigashi@andrew.cmu.edu

**Abstract:** Community-based afterschool programs are valuable spaces for researchers to codesign technologies with direct relevance to local communities. However, afterschool programs differ in resources available, culture, and student demographics in ways that may impact the efficacy of the codesign process and outcome. We ran a series of multi-week educational game codesign workshops across five programs over twenty weeks and found notable differences, despite deploying the same protocol. Our findings characterize three types of programs: *Safe Havens, Recreation Centers,* and *Homework Helpers.* We note major differences in students' patterns of participation directly influenced by each program's culture and expectations for equitable partnerships and introduce Comparative Design-Based Research (cDBR) as a beneficial lens for codesign.

## Introduction and Background

Design-Based Research (DBR) is a methodology for producing generalizable knowledge about complex systems while also creating products that operate effectively within those systems (Barab, 2006). Borrowing the iterative logic of engineering design, DBR involves a cycle of building, testing, and improving prototypes in the context where they will be used. Due to the situated nature of the work, the products of DBR are influenced by the environments in which they are built. The interplay of context and creation allows DBR to not only facilitate the design of effective and engaging artifacts but also yield insights about situated environments, valuable to the Human-Computer Interaction (HCI) and Learning Science research communities.

However, a challenge remains in understanding specific constraints and unique strengths of these insights given they are derived from a highly contextualized process. When conducting DBR in settings where researchers are not members of the same social or cultural group as the intended users, participatory design methods, such as collaborative design (codesign) are often employed (Guha et al., 2013). Acknowledging that researchers' expertise does not extend to cultural and environmental norms in the settings where the products will be used, participatory design invites members of the community into the design process to provide insight into how products or programs could be designed for better cultural fit. However, it remains unclear how similar a new community would need to be for the insights to generalize, or perhaps how the design decisions could be manipulated for a differing community, given a lack of contextual information — i.e., more exact constraints, potentially limiting the findings.

Afterschool centers are a great environment to conduct adult-child DBR related to transformational games, as formal school schedules may be too rigid to allow for the unstructured play required to create highly engaging games. They cater to students' varied interests including sports, games, arts, STEM, etc. and routinely partner with external organizations to offer programs that students are highly interested in. Recently, afterschool STEM programs have been greatly promoted. Demand for such centers exponentially increased during the pandemic (Afterschool Alliance, 2022) and has remained strong even as schools have re-opened. Centers are expected to provide safe, supportive, adult-supervised environments, as well as academic, personal, social, and recreational development. Many providers are interested in expanding (or introducing) STEM program offerings, but not all STEM resources have proven successful in all environments. In fact, the multitude of virtues ascribed to afterschool centers creates an ever-growing list of expectations that partners are expected to meet, increasing pressure on providers to constantly change their programs to incorporate new, in-demand elements.

Codesign encapsulates multiple approaches, serving as a popular methodology to co-create experiences, products, and systems with users, stakeholders, and community members (Zamenopoulos & Alexiou, 2018). Codesign supports developing ideas, better designs for user and stakeholder needs, creating relationships between designers and groups of people, and quality results (Steen et al., 2011). Power sharing and shifting emerges from codesign, as stakeholders are given a pathway to provide insight and share needs (Westin & Salén, 2019). Often, codesign has been done with adult participants, however, as children utilize technology and are stakeholders in their own STEM educational experiences, it is necessary to include youth in codesign as well. Children's own insight is critical to designing effective experiences (McNally et al., 2016; Yip et al., 2013), since they are experts of their own interests. Their expertise is especially beneficial when the target goal is to create STEM educational products, such as transformational games (games developed with the intention of changing players in a specific way that transfers outside of and persists beyond the game) for diverse groups (Culyba, 2018).

Walsh et al., (2013) describe codesign as focusing on the design partners (their experience, needs for accommodations), design goal (design space, maturity of design), and design technique (cost, portability, physical interaction, and technology). However, when working with children, it is important that the techniques and framework for design fit the population and creativity is harnessed (Read et al., 2002). The needs for children from under-resourced or marginalized communities to become codesign partners may also be different and require a critical eye towards justice, diversity, and accessibility from codesigning researchers, e.g., drawing on critical race theory and intersectionality to better center diverse child codesigners (Ogbonnaya-Ogburu et al., 2020). Effectively involving diverse children in codesign can allow a range of participants to contribute valuable design insights. In particular, we aim to fill a gap in understanding how to effectively codesign with a range of children across different kinds of afterschool learning environments through a comparative approach.

In this study, we employ "comparative design-based research" (cDBR) to examine differences in barriers to implementation of STEM programs across five culturally and economically diverse afterschool centers. Our DBR program (the same program run across multiple sites) sought to create a robotics programming video game that would be engaging and practical within each cohort's environment. Each site's implementation spanned approximately 20 weeks and focused on codesigning a game that the youth in each cohort would like to – and be able to – play in their afterschool center. As expected, the codesign process produced inferences about the design of STEM games, however the cDBR approach also highlighted how different sites with communities of youth across different racial, age, and socioeconomic lines resulted in substantially different codesign experiences across populations. For the scope of this paper, we focus on the partnerships, codesign processes, and experiences to highlight methodology. Findings related to the game artifacts will be reported in a future publication.

In this paper, we argue that leveraging a cDBR approach provides insight and value to involving diverse youth in codesign work. We will demonstrate that beyond the conventional considerations that researchers take into account when designing research programs (such as racial and socioeconomic composition), additional critical factors to the conduct of participatory design-based research emerge from consideration of the internally aligned sets of goals, scheduling, and youth participation norms that follow from a program's structural character. The structural forms of different afterschool settings can be characterized by a common set of archetypal identities.

## Context and Method

Our program took place in a group of afterschool programs in and around a mid-sized city in the Mid-Atlantic region of the United States during the 2021-2022 academic year. Our goal was to collaboratively create video games with different afterschool sites where players program robot partners to accomplish goals together. The game was intended as an educational product, as well as a pedagogical exercise to foster students' identity as designers, increase their programming knowledge, and robotics domain knowledge.

We ran a 20-week program focusing on the codesign of game characters and narratives, block programming instruction, and game testing and iteration. In weeks 1-8, students were informed that they were partnering with our research team to create a video game where they collaborated with a robot by programming it to accomplish their in-game goals. Using different codesign instruments and activities, we explored different game narratives, co-created game characters and settings, negotiated game mechanics etc. The next 8 weeks were focused on block programming instruction to equip students with the skills needed to program their robots. This time also allowed our software development team to create a prototype version of the game. During the final 4 weeks, we iteratively tested the game with students. Students tested and critiqued prototype and beta versions of the game, often with us modifying the game elements, narratives, and mechanics between versions. Each codesign session was one hour long and consisted of snacking and icebreakers, scheduled codesign activities, and students playing diverse-genre video games from a curated selection. We obtained written consent from students' parents/guardians and our research was approved by our university's Institutional Review Board (IRB).

Each session was attended by 2-4 researchers (depending on the number of students), with one being a dedicated notetaker. One researcher (second author) was the main lead for most sessions at each site, while the others had supporting roles (taking notes, answering questions, passing out materials, etc.). Site staff members sometimes joined our sessions and assisted with the facilitation. The program was structured for 6-8 students but adjusted to accommodate much larger groups as necessary. After each session, the researchers met to discuss the data gathered and clarify any areas of confusion. These meetings were recorded and transcribed as well. All team members attended a weekly analysis meeting to review session interactions from the different network sites, design and refine planned session activities, and reflect on the types of program adjustments we needed to make to better serve each afterschool center. We recorded these analysis meetings and analyzed them as part of the data.

**Table 1**
*Afterschool club information (collected from interviews) showing total students, age range, and demographics.*

| Club name | Total # students | Age range | Racial demographics description |
| --- | --- | --- | --- |
| Green Hill | 30 (10 F, 20 M) | 5-14 | Mostly Black, some White, some Hispanic |
| Central Rise | 12 (12 F) | 11-14 | Predominantly Black |
| Sunny Pond | 19 (6 F, 13 M) | 7-12 | Predominantly White |
| Clear Bridge | 15 (3 F, 12 M) | 7-10 | Predominantly White, some Hispanic, some multiracial |
| West Creek | 48 (21 F, 27 M) | 5-12 | Diverse (including White, Black, Hispanic, Middle Eastern, multiracial) |

We worked with 5 afterschool clubs (Table 1): Green Hill (former industrial low-income town with mostly Black families), Central Rise (low-income neighborhood in city center with mostly Black families), Clear Bridge (rural town with mostly White families), Sunny Pond (suburban town with mostly White families), and West Creek (diverse suburban town). Site names are pseudonyms. All programs took place in-person, except for at Sunny Pond, which was run mostly remotely due to COVID regulations with the help of a staff member onsite.

We have several sources of data: Staff interviews, observation notes, researcher reflection meetings, student surveys about technology and programming experience, and codesign data (including design artifacts, playtesting session observations, etc.). Staff interviews (N =10; 5 F, 5 M; 5 White, 3 Black, 1 South Asian, and 1 Hispanic) consisted of 8 who were dedicated to their individual centers, while 2 were administrators who served as STEM coordinators for all afterschool centers in the network. Each interview, focused on the culture of the programs, their available resources, and values, lasted one hour and was recorded for analysis. Participants were not financially compensated but were informed that participation would allow us to better cater to their students.

We conducted a thematic analysis on the different data sources (Braun & Clarke, 2012) to understand and characterize how organizational goals, resources, activity provision norms, and student participation criteria impacted the effectiveness of our codesign program and required different adaptations. We reviewed the interviews first, then session observations, and meeting notes for data that provided evidence for each identified theme. Given the focus of the interviews, preliminary themes highlighted aspects of program culture. These themes were then corroborated and iterated based on analysis of the other data sources. After this process was completed, the entire team discussed each theme extensively to clarify areas of confusion. Where necessary, the team watched session videos as a group where multiple perspectives were needed to unpack the interactions. We triangulated our findings with student-generated artifacts to ensure that all evidence was mutually supportive. Our team consisted of a range of researchers coming from computing, learning science, robotics, and game design academic backgrounds, as well as a variety of economic, racial, and cultural backgrounds (including East Asian, Black, and White, with American and African cultural backgrounds). Researchers often had multiple roles on the teams, including being facilitators, programming and building the games, and designing the games.

## Findings

We identified three "program archetypes" across the five sites with characteristically different goals, schedules, and activity participation norms, which in combination had a dramatic impact on the conduct of afterschool programs (including our DBR activity, which was shaped as a STEM program). These categories are not meant to be exhaustive, or comprehensive for describing all kinds of afterschool program partners. Researchers may encounter partners who do not have any of the factors described, or more likely, partners who have characteristics that span across these archetypes. These characterizations provide insight into how different communities can be contrasted using one design program protocol. We find differences across sites by using a comparative approach.

## Program archetypes: *Safe Havens*, *Homework Helpers*, and *Recreation Centers*

### *Safe Havens*
We categorized Green Hill and Central Rise as *Safe Havens*. We interviewed three directors from both programs (2 Black men, and 1 Black woman). These afterschool programs are located in high-poverty neighborhoods with mostly Black and Brown students attending low-performing schools. The primary purpose of these sites was to provide a safe space for children to stay after school. Unlike our other sites, all the students in these clubs qualified for financial assistance and therefore attended the program for free. They have strong partnerships with local

schools – Green Hill is physically located in the local elementary school, and Central Rise has liaisons who work in the schools they serve. The Central Rise program director mentioned, *"We help with education… education by far, because not only are we on them when they get here or just when they're not here, we keep in contact with 'em. We have a liaison up at the school. Okay? Who's also checking on them, their mannerism …"*

Staff members in *Safe Havens* did not only serve as enrichment coordinators but also as parent and guardian figures. For example, the program director at Green Hill shared that he regularly raised funds from different organizations to ensure the students had clothes and school supplies – on several occasions he invited hair braiders and barbers to the afterschool programs to ensure students had presentable hairstyles. In both programs, members sometimes transported students in their personal vehicles, cooked meals for them, and advocated for them in schools and with their families. Taking on this parental role came easier for all three directors as they could personally identify with the students – they all shared that they grew up in similar neighborhoods, faced similar struggles, and attended afterschool programs with adults that made a difference.

In addition to providing a safe space and serving as parental figures to students, these administrators shared that exposing their students to more opportunities regardless of their interests was important to them. Central Rise had organized international trips to Ghana and Jamaica in the past to expose their students to people from other cultures, and Green Hill was constantly looking for more exposure opportunities.

> *"It's just like abilities to see things. Whether it be resources or supplies or whatever it may be ... So, for me, the idea is like the exposure. I remember like going to a tennis camp… I didn't care about that tennis camp, not one bit, but now that I'm 31 years old, I can see someone playing tennis on TV, and I can appreciate that because at a young age, I was kind of exposed. I don't think that the exposure was so much about me like becoming Serena's [Williams] little homie. It wasn't about that. It was about me experiencing it ... We're going on a walking trail tomorrow ... these city kids have no idea what a walking trail looks like."* — Green Hill program director

Although Green Hill is part of a larger afterschool program network, they are physically located at least 30 minutes from the city center and are largely inaccessible by bus. Therefore, they do not benefit from the STEM programming and staff available to the other flagship afterschool sites located in the city. They also have very limited computing infrastructure; when we first started working with them, there were only three computers available to both students and staff. Due to limited opportunities for STEM programming, administrators welcome lasting external programs to accommodate many students with minimal infrastructure and staffing requirements.

Central Rise is in the city downtown and has better access to STEM programs and researchers from local universities, non-profit organizations, and technology companies. However, they also struggle with maintaining consistent STEM programming, as they didn't have the staff to provide instruction in house. Throughout the COVID-19 pandemic lockdown, our research was the only external program available to them. Although they had some donated computers, they were all severely outdated, and we had to provide technical support while working with them including replacing missing peripherals, operating system updates, upgrading memory, etc.

### *Homework Helpers*

We categorized Clear Bridge and Sunny Pond as *Homework Helpers*. We interviewed three directors from both programs (one white woman, one white man, and one Indian-descent man). Clear Bridge is located in a rural town where white working-middle class families make up 95% of the population. Sunny Pond, on the other hand, is in a small town in close proximity to other afterschool programs in the network. Working class white families make up the majority of the population. Families in both clubs primarily signed up their children to get homework help from the coordinators. The program director of Clear Bridge described, *"We are homework help. That's a main one… Parents come for the after-school homework help. They like that. They like that we have structure here that we do the academic things… It's the ability because [parents] don't understand the math … So, when they get home, they're not having to worry about rushing around, getting homework done. They can actually spend time with their kids."* Focusing on academics makes *Homework Helper*s conducive to running structured programs.

> *"Students [at* Homework Helper *programs] will buy in enough to do what they're told, because someone told them to do it, whether or not they [fully] buy in ... Most of the students are like, 'Okay, I will do what I'm told. I will sit here. We may not be passionate about it.' So, like when I picture [other programs in the area], we have the crazy, like, nothing goes smoothly, ever. All the kids are all over the place, but there's a lot of kids that are passionate and like excited about it… [At Sunny Pond and Clear Bridge], there's a little bit of crazy, but a lot more kids with less passion."* — Network-level STEM director

Since parents entrusted the administrators at the *Homework Helper* sites with the responsibility of overseeing all afterschool academic activities, they did not get involved in what activities their children participated in. Administrators had the responsibility of pre-selecting students who attended different programs based on their perceptions of student ability and interest. At Clear Bridge, administers gave us a list of students whom they felt were best suited for our program. While most of the selected students had prior programming experience and game design interests, several other students who could have benefited from the program were left out. On one occasion, we had the opportunity to share our program information with all students who attended the club on a particular day, and seven new students came to us informing us that they were interested in joining. Sunny Pond had a more flexible signup structure. Like Clear Bridge, the administrator pre-selected the students who they felt were best suited for our program, but they often reevaluated other students who visited the club.

Both afterschool sites had adequate technology infrastructure. Clear Bridge had a computer lab with touchscreen desktop computers. There was a general space for homework, and a makerspace with more computers and programmable robots. Sunny Pond did not have a dedicated computer lab, but they had laptops available for each student and several classroom-style spaces that were used for different programs. Both sites had an in-house STEM champion. The program director at Clear Bridge had taught several STEM programs in the past and was regularly soliciting partnerships for opportunities. Sunny Pond had a dedicated STEM programming staff member.

### *Recreation Centers*

We categorized West Creek as a *Recreation Center* and interviewed two program directors (one Black man, and one Hispanic man) for this study. Prior to their serving as directors at West Creek, they worked as program directors at Green Hill and another local center not in this study, so they could reflect on their experiences working in the different settings. One program director mentioned, *"Your role might have to shift depending on the neighborhood and the kids that you're working with. [The other director] and I have a bunch of cartoon stuff around our office, and I think that exemplifies the vibe that we have here as kind of a, 'Hey, let's hang out buddy' type of thing ... I don't think they look at us as father figure sort of thing ... [We're] one of the, one of the homies."*

When describing the purpose of their club, the program director of West Creek explained the culture of the club centering fun and entertainment: *"Kids either come here for sports, or they come here to do the fun kind of hangout things that are happening... The idea of the club is fun. ... You know, like when you're in school ... if they make a game out of learning, that's exciting. If you go to the club, and they make a game out of learning, and [the children may think] 'Whoa, whoa, what is this? Oh, we don't want to do this.'"*

In addition to a dedicated computer lab, West Creek has a dedicated gaming room with several video game consoles, board games, and billiards for teenagers. They have an open lobby space with arcade style video games and consoles, billiards, and table tennis. They also have two dedicated gyms where multiple sports programs were conducted, an outdoor playground, an art room, and a dedicated maker space. While administrators recommended and encouraged students to join certain activities, the students have the agency to leave and join any other activities in the building at will. We observed several incidents where students joined the one activity but were informed by their friends that something cooler was happening elsewhere encouraging them to leave.

## Norms around participation in codesign program

To illustrate the utility of our cDRB approach, we describe one main theme from the data, characterized by differences between archetypes touching on culture, values, resources, and demographics. In particular, we saw that the norms around and barriers to consistent participation dictated the types of data that could be collected and mediated how learners could engage with our program. Table 2 shows the attendance patterns at different clubs.

Attendance patterns at Sunny Pond and Clear Bridge were most consistent with our prior expectations for our program's participation. Most students who signed up attended most sessions as long as they were in the building. Although Sunny Pond and Clear Bridge had students attend less than 40% of all sessions on the average, the main reason for this was that these clubs were most sensitive to set of fee increases implemented by afterschool network partway through our program, as most children came from working to middle class families and did not qualify for financial assistance. There was a program wide-fee increase from $90 to $110 per week which caused some parents to pull their children from the club. For students who remained, attendance was higher. We made the fewest modifications to our original program design with *Homework Helpers*. Students attended consistently enough, provided codesign input, and were invested in the game output. There was less opportunity for identity transformation and learning gains compared to *Safe Havens*, and less diversity in game feedback compared to *Recreation Centers*, but their predictability was most beneficial for the original codesign process and product.

Although they provided the most consistent student attendance, conducting research in *Safe Havens* was quite challenging. Despite having only 12 students, attendance was highly consistent at Central Rise. Most students at Green Hill attended 43% of all sessions — however, this was further reduced by months of including

**Table 2**

*Participation information and archetype. Clear Bridge and Sunny Pond experienced a tuition hike that impacted the number of families that attended the afterschool program. *The second numbers listed for these afterschool clubs are the average for students who remained in the program despite the tuition increase.*

| Club name | Archetype | Avg. % of total sessions each student completed | Avg. # of repeat students per session |
|---|---|---|---|
| Green Hill | *Safe Haven* | 43% | 11.45 |
| Central Rise | *Safe Haven* | 71% | 7.45 |
| Sunny Pond | *Homework Helper* | 31% (*53%) | 5.16 |
| Clear Bridge | *Homework Helper* | 39% (*50%) | 4.8 |
| West Creek | *Recreation center* | 13% | 3.95 |

some students and excluding others, as we tried to maintain a maximum of 10 students in our program to avoid crowding in small spaces and support meaningful codesign interactions. Attendance was more consistent when we could provide enough research staff to accommodate all students. Staff members insisted that equitable provision of programs to their students involved mandatory participation. We tried different approaches to account for the number of students including taking half of the students on one day and coming the next day to repeat the program for the other half. That too, was not sufficient as staff members could not guarantee student attendance, so we had to cater to all students without additional time available. We also needed to spend more time teaching prerequisite skills. Students at Green Hill had mastered fewer arithmetic, reading, and STEM competencies compared to their middle- and upper-class counterparts. The mandatory participation requirement also meant that we had varied student interest in our program and had to work with students who did not want to be there. Given the low prior STEM knowledge and mandatory attendance requirements, students in *Safe Havens* showed the highest potential for STEM identity transformations and learning gains. Long term programs that make strong connections between STEM and other domains provide consistent attendance by research partners and expose students to different kinds of professionals seemed especially useful for this archetype of afterschool program.

In *Recreation Centers*, we had the most inconsistent attendance of all our programs. West Creek students attended an average of 13% of all sessions, and approximately 4 students were repeats compared to the 48 students we encountered in totality. Students in *Recreation Centers* shuffled between multiple spaces within the same time block looking for activities that seemed the most exciting. When students opted to play outside, play video games, or play sports in the gym, it was unlikely that they would stop by our program. When we had students in the room, there was high interest, but the attitude was all about play. One potential strategy to increase student participation was to communicate our program goals, milestones, and output directly to parents so they could encourage their students to attend. However, we did not have direct access to parents and lacked the consistency in output materials per child to make that strategy feasible during our time there. Given these attendance patterns, it was impossible to conduct pre- and post-testing, observe long term identity transformations, or conduct programs not centered around play or fun. Students in *Recreation Centers* provided inconsistent input in the codesign process, but their input was particularly valuable when testing new iterations of our video games. Because they did not have the same emotional ties and investments as groups who consistently informed the design of the game, their feedback was critical for understanding how each game might be understood by fresh students from other demographics.

## Discussion

The outcomes of Design-Based Research are typically twofold: contextually appropriate products, and reflections on the context-product interactions that make them so. Typically, these insights are powerful and authentic but also inherently specific to a single context. In this paper, we presented findings from our effort to deploy a STEM codesign program across multiple settings. True to its DBR roots, this work produced data and analytic leverage from which to reason about the context (afterschool programs) in which the work was taking place, surfacing characteristics of afterschool programs that affect such programs. In particular, because our STEM codesign program served as a data collection instrument, we were able to identify numerous factors relevant to the conduct of design-based research in these settings. For instance, inconsistent student attendance at programs could result from several different forces – in programs with high levels of free choice, it could be the result of our activities being unappealing; in other programs attendance was constrained by level of trust; in others, it was beyond our control entirely. In other cases, the ways in which norms affected codesign participation could be quite subtle. For example, *Recreation Centers* maintained a culture of play and value of fun, making it trickier to direct youths' attention toward creation of more serious artifacts, especially when other activities like sports were offered at the same time. Vitally, it is the fact that we were able to observe the same STEM codesign offering playing out in these different ways between sites that allow us to be confident of the different mechanisms by which this occurred. Our findings corroborate prior literature characterizing types of out-of-school programs, relating to e.g., homework assistance (Cosden et al., 2001) and its benefits to learners. We extend such work by noting that programs in fact organize themselves quite substantially around core services they offer, such as homework help, creating differences across the culture of the organization, impacting many activities, including codesign.

### Educational and methodological implications

Regarding implications for informal education research, by focusing on the differences between common patterns of program structure we found that codesign-critical program participation norms and routines could be highlighted using archetypes based on administrators' characterizations of programs. Comparing the alignment of administrator characterization to observed program routines across sites allowed us to see if that administrators' visions seemed to be the genesis of scheduling and attendance norms across sites, or at least that administrators were reliably aware of such alignments. While racial, socioeconomic, and broader structural factors are likely at play and cannot be separated from our findings given the nature of our specific sites, the archetypal descriptors provided micro-level insight into the site routines that determine whether a codesign session will be productive.

This work also yields methodological implications. The findings we present here are primarily triangulated from observed *differences* between *sites*. However, between-sites comparisons are not the only analytic leverage provided by cDBR. We have also seen promising results emerging from exploration of *commonalities* between the *products* resulting from codesign at each site (Higashi et al., 2022). We may thus hypothesize that, at a minimum, a cDBR approach could produce inference based on something akin to a 2x2 matrix of either *similarities* or *differences,* between *sites* or the *products* they generate – the present study, for instance, focused on *differences between sites*. Yet the matrix-cell approach ignores the inextricable link between contexts, processes, and products in DBR. Indeed, some of cDBR's greatest inferential power may come from examining combinations of these factors. We can readily imagine, for instance, a situation where certain product design features emerge across multiple different sites for different, yet independently sensible reasons. *Product similarities* despite *site differences* in design rationale would provide provocative evidence of *convergence*, supporting the existence of important invariants around DBR. Such insights remain notional now but point to a powerful potential for comparative methods in design-based research, which we will pursue in our future work.

We recommend that researchers who intend to use cDBR keep a few points in mind. A diverse range of programs will provide the greatest leverage for triangulation. Differences and therefore findings across sites may be more robust with greater diversity across multiple axes. Larger differences in site context will entail a wider array of needs across teams and communities, impacting the overall nature of the work and increasing the need for flexibility in the codesign protocol. Preparation for this may include, for example, ensuring the ability to call in extra research team members to support facilitation onsite for certain collaborations.

### Limitations

There are some threats to validity. First, our evidence is drawn from a single codesign effort conducted across 5 sites in a single, geographically contiguous network of afterschool programs. Therefore, while our argument establishes an existence proof of the value of cDBR, determining the full extent and scalability of cDBR methods is left for future research. Similarly, we acknowledge that while we intentionally employed a diverse team of researchers and took an iterative process in our analyses, there are inescapable possibilities of bias when analyzing data and the inherent advantages and disadvantages of researcher-as-instrument in qualitative work. Additionally, a programming-based co-robotic game may result in prior knowledge highly influencing participation, and different insights might be uncovered for different types of products in other domains. Finally, by nature, some of our findings about suitability and responsive adaptation of DBR instruments to different contexts emerged from failure. Thus, while analyses of these breakdowns allow us to observe the failure mode, conditions under which issues occurred, and immediate causes, it is impossible to directly observe the counterfactual, i.e., what the program would have looked like had we incorporated needed adaptations in advance. We intentionally scope this work around the methods, showing the potential of this deployment of cDBR across sites. Broader results, including additional themes related to the design artifacts, including the games, will be reported in the future.

## Conclusion

A cDBR approach, employing the same codesign protocol and processes in intentionally different settings, yields important insights about interactions between protocols and settings, and therefore serves to highlight particular strengths or limitations in the codesign findings. Without the comparative multi-site aspect, we would have not had a point of comparison from which to argue that interesting codesign process decisions stemmed from an interaction with systemic local conditions. While the focus of this paper has been to highlight the potential for cDBR through emphasizing differences between sites, future work will explore the *persistent* or *convergent* design findings that can also be made apparent by the approach.

## Acknowledgements

This work was funded by NSF AISL 1906753. We thank the children who we worked with, as well as the school directors and staff who helped support the programs and work that we carried out.